# Textile-based conformable and breathable ultrasound imaging probe

Takumi Noda, Seiichi Takamatsu, Michitaka Yamamoto, Naoto Tomita, Toshihiro Itoh,
Takashi Azuma, Ichiro Sakuma, and Naoki Tomii

*Abstract—* **Daily monitoring of internal tissues with conformable and breathable ultrasound (US) imaging probes is promising for early detection of diseases. In recent years, textile substrates are widely used for wearable devices since they satisfy both conformability and breathability. However, it is not currently possible to use textile substrates for US probes due to the reflection or attenuation of US waves at the air gaps in the textiles. In this paper, we fabricated a conformable and breathable US imaging probe by sandwiching the US elements between two woven polyester textiles on which copper electrodes were formed through electroless plating. The air gaps between the fibers at the electrode parts were filled with copper, allowing for high penetration of US waves. On the other hand, the non-electrode parts retain air gaps, leading to high breathability. The fabricated textile-based probe showed low flexural rigidity (0.066 × $10^{-4}$ N·m²/m) and high air permeability (11.7 cm³/cm²·s). Human neck imaging demonstrated the ability of the probe to monitor the pulsation of the common carotid artery and change in the internal jugular vein diameter, which lead to the early detection of health issues such as arteriosclerosis and dehydration.**

## I. INTRODUCTION

WEARABLE devices that perform the continuous and noninvasive monitoring of health conditions have gained attention in recent years [1]–[3]. In conventional medicine, diagnoses and treatments are usually made after symptoms appear [4]. As a result, the diseases often progress to the severe stage. Daily monitoring of health conditions using wearable devices is promising for early detection of diseases [5]. To ensure effective daily monitoring, it is crucial that the device has high flexibility and breathability. Low-flexibility prevents close contact between the device and the skin, resulting in reduced signal quality [6]. Low-breathability prevents sweat excretion of the skin, leading to inflammation [7], [8].

In recent years, wearable devices fabricated on electronic textile (e-textile) substrates have gained significant interest because textiles have high flexibility and breathability. Various wearable health monitoring devices have been developed using e-textile substrates. Paul *et al.* fabricated a wearable ECG monitoring device by printing the silver paste on a woven textile [9]. Husain *et al.* fabricated a wearable temperature-sensing device by laying-in a metallic wire into a knitted structure [10]. Although these textile-based wearable devices can monitor some vital signs during daily activities, they cannot perform the imaging of internal tissues. Internal tissue imaging is crucial for the early detection of abnormalities in organ morphology and dynamics, as these abnormalities usually do not appear in vital signs until they have progressed to the extent of causing functional issues. For example, valvular heart diseases are typically difficult to be detected from ECG or blood pressure in their early stages [11].

Ultrasound (US) imaging is one of the methods to visualize morphology and dynamics of internal tissues in a minimally invasive manner. Conventional US imaging is performed through the following processes. First, a US probe is pressed against the body to ensure close contact. Close contact between the probe and the body is necessary for US imaging because more than 99.9% of the US wave is reflected at the interface when there is an air between them [12]. Then, the probe acquires US signals by transmitting US waves into the body and receiving backscattered waves from tissues. Finally, a US image of inside of the body is reconstructed from the US signals. Development of wearable US imaging devices will improve the accuracy of disease detection in daily health monitoring.

There have been various studies to develop wearable US probes. In some studies, conventional hand-held US probes were made wearable using fixators [13]–[15]. However, fixators may harm patients by pressing the probes against the body with great force to ensure close contact. A thin and rigid US probe was recently developed that can be attached to the skin without pressing [16]. The surface of this probe was coated with hydrogel and bio-adhesive layers, which provide adhesion to the skin. In other studies, flexible US probes have been developed, which have US elements arranged on thin and flexible sheets [17], [18]. Since flexible probes deform according to the geometry of the body surface, they can be attached to the skin with small mechanical stress even when the body is moving, or the body surface is uneven. Elloian *et al.* developed a flexible US probe by integrating

US elements on a polyimide film substrate [19]. Hu *et al.* developed a flexible and stretchable US probe using elastomer substrates [20]. The developed flexible US probes succeeded in measuring the brightness mode (B-mode) imaging, Doppler flow imaging, and elastography [21], [22].

Although the aforementioned flexible probes adhere to the skin with small mechanical stress, they are still less comfortable when used for long-term monitoring. Specifically, the substrates of these probes are plastic or elastomer films, which are not breathable and may cause inflammation. It is desirable to fabricate flexible probes on breathable substrates such as textiles. However, a flexible probe with textile substrates has not been made because the US waves transmitted from the elements cannot penetrate the body through textiles (Fig. S1). Two main factors prevent the US waves from penetrating the textiles. The first factor is that acoustic energy is converted into thermal energy due to friction between the fibers [23]. The second factor is that textiles are porous and contain air within their structure. The acoustic impedance of air (~400 Rayls) [24] is significantly lower than that of the textile materials such as polyester (~3.0 MRayls) [25], and the US waves are reflected at their boundaries due to the acoustic impedance mismatch [26]. When the US wave propagates from polyester to air, the transmission factor is only $3.0 \times 10^{-5}$. The transmission factor ($TF$) when US wave propagates from a medium with the acoustic impedance of $Z_1$ to a medium with the acoustic impedance of $Z_2$ is calculated by the following equation [27]:

$$TF = \frac{2Z_2}{Z_1 + Z_2} \qquad (1)$$

Here, we show conformable and breathable ultrasound imaging probe fabricated on partially filled textiles for internal body health monitoring. The textile substrates were fabricated by partially forming copper electrodes and wirings on the thin woven polyester textiles by electroless plating. As shown in Fig. S3, the air gaps between the polyester fibers at the electrode parts are filled with copper, allowing for high penetration of US waves. Conversely, the non-electrode parts have air gaps, resulting in high breathability. The textile-based probe was fabricated by sandwiching the US elements between two e-textiles. The signal electrodes for the elements and the common ground electrode are printed on the top and bottom textiles, respectively. The textile-based probe showed low flexural rigidity ($0.066 \times 10^{-4}$ N·m$^2$/m), high breathability (11.7 cm$^3$/cm$^2$·s), and high stability against large and repeated deformations. The lateral and axial imaging resolutions of the textile-based probe at 10 mm depth were 0.31 and 0.64 mm, respectively. The textile-based probe clearly visualized simulated blood vessels with a diameter of 6 mm in longitudinal and transverse planes. The results of human neck US imaging showed that the pulsation of the common carotid artery and the change in the diameter of the internal jugular vein can be monitored by the textile-based probe.

## II. RESULTS

### A. Device fabrication

Fig. 1a shows the optical image of the textile-based probe attached to the clothing. The probe was fabricated by sandwiching the US elements between two e-textiles on which the electrodes and wirings were formed (Fig. 1b). Fig. 1c shows the cross-sectional image of the probe taken by scanning electron microscopy (SEM). The magnified views of the junctions between the e-textiles and the US element are shown in Fig. S2. As shown in Figs. 1d and 1e, the probe has high flexibility and breathability.

The US elements were made from 1-3 piezoelectric composites (Fuji Ceramics Co., Ltd.), which were fabricated by dicing the lead zirconate titanate (PZT) and filling the kerfs with epoxy (Fig. 1c). Piezoelectric composites have better acoustic impedance matching with human tissues compared to single-phase piezoelectric ceramics [28]. The top and bottom surfaces of the 1-3 piezoelectric composite were coated with Au by sputtering. The elements were cut from a wafer using a blade dicer (DAD3650, DISCO Co.). The elements' width, length, and thickness were 0.50, 5.00, and 0.29 mm, respectively. The thickness of the elements was determined so that the resonant frequency was 5 MHz. The number of elements was 16, and they were arranged on the e-textiles with a pitch of 1 mm.

The e-textiles were fabricated by forming the electrodes and wirings on woven polyester textiles (Seiren Co., Ltd.) by electroless copper plating. The thickness of the textiles was 0.09 mm. Fig. S3 shows the SEM image of the e-textile. On the top textile in Fig. 1b, signal electrodes of the elements and wirings leading to them were formed. On the bottom textile, a common ground electrode for all the elements and a wiring leading to it were formed. In the electroless plating, copper is deposited onto the polyester fibers not only on the textile surface but also within the textile. This process fills the air gaps between polyester fibers and enables the US waves to pass through the bottom textile with high penetration (Fig. S3). Furthermore, because the acoustic impedance of polyester (~3.0 MRayls) [25] is between that of piezoelectric composite (~8.5 MRayls) [28] and human tissue (~1.6 MRayls) [29], the e-textile enhances the US wave transmission factor from 0.10 to 0.32. This eliminates the need for an acoustic matching layer, which could potentially decrease the flexibility and breathability of the probe [26].

The e-textiles and the US elements were bonded using low temperature solder (L20-145HF, Senju Metal Industry Co., Ltd.) at 160 °C for 10 s. Bonding was performed under pressure by placing the materials between a magnet heater (Silicone Rubber Heater, O&M Heater) and a steel plate. As shown in Figs. 1c and S2, the gaps between the elements and the e-textiles were filled with solder, resulting in improved US permeability. Even though each textile possesses high flexibility, the probe's flexibility was reduced because it was constructed with two textiles joined through the US elements. In order to address the reduced flexibility resulting from the structure, cut-outs were made between the successive electrodes of the top textile (Fig. 1b). Furthermore, because the probe doesn't adhere to the skin as it is, an adhesive hydrogel sheet was attached to the element array section of the bottom e-textile for US imaging (Fig. 1b). Attaching the hydrogel sheet has little effect on US wave propagations because the acoustic impedance of the hydrogel is almost same as human tissues (~1.6 MRayls) [30].

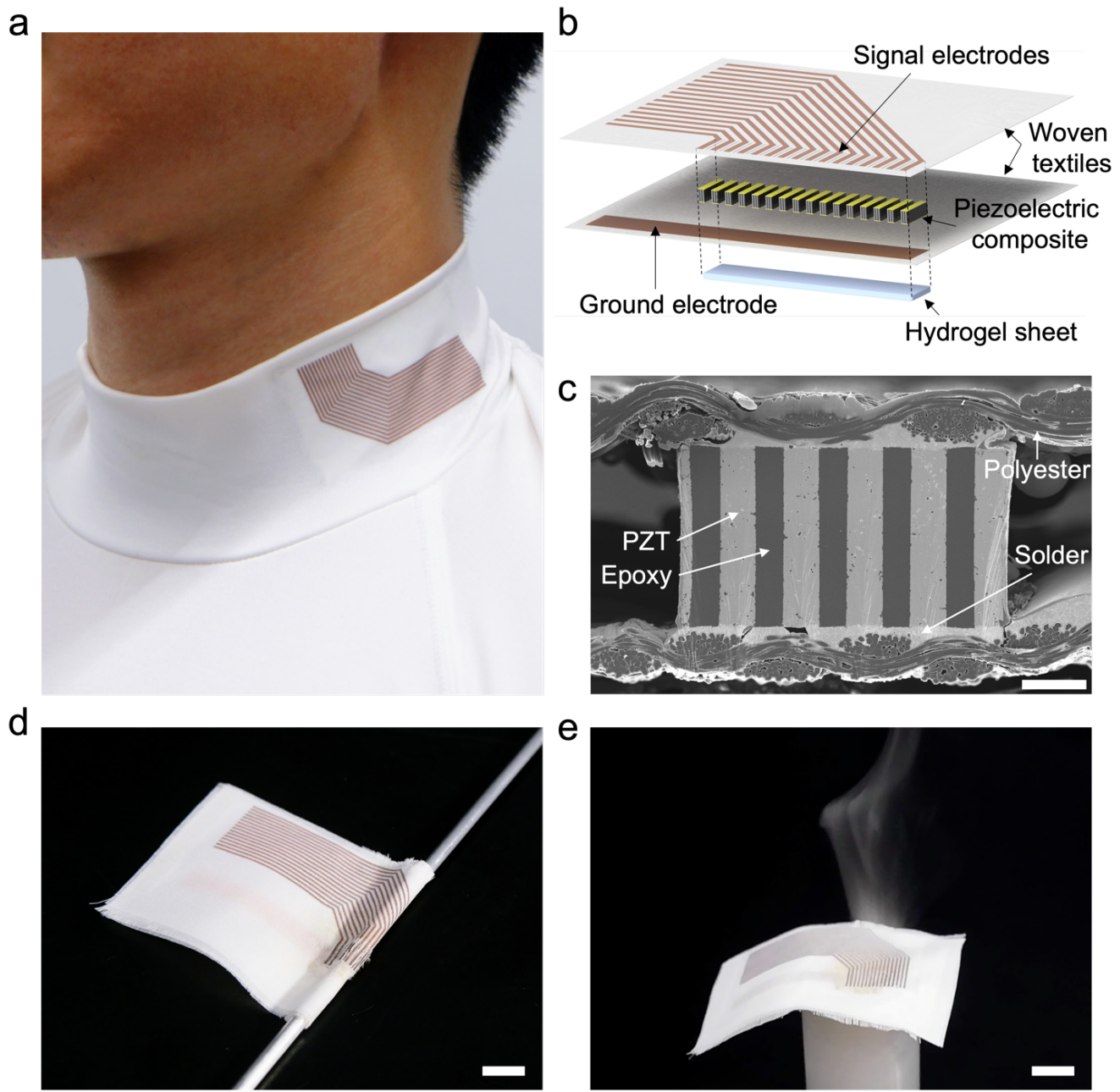

Figure 1. A textile-based ultrasound imaging probe. (a) Optical image of the textile-based probe attached to clothing. (b) Schematics of the textile-based probe. The probe was fabricated by sandwiching the ultrasound elements between the e-textiles. (c) Cross-section image of the fabricated textile-based probe taken by scanning electron microscopy. The scale bar is 100 $\mu$m. (c) (d) Optical images showing the flexibility and the breathability of the textile-based probe, respectively. The scale bar is 10 mm.

In order to compare the performance of the textile-based probe with the probes using plastic and elastomer film substrates, we also fabricated probes with polyimide and polydimethylsiloxane (PDMS) films using a similar process. The thicknesses of the polyimide and PDMS films were 0.06 mm and 0.10 mm, respectively. Fig. S4 shows the polyimide-based and the PDMS-based probes.

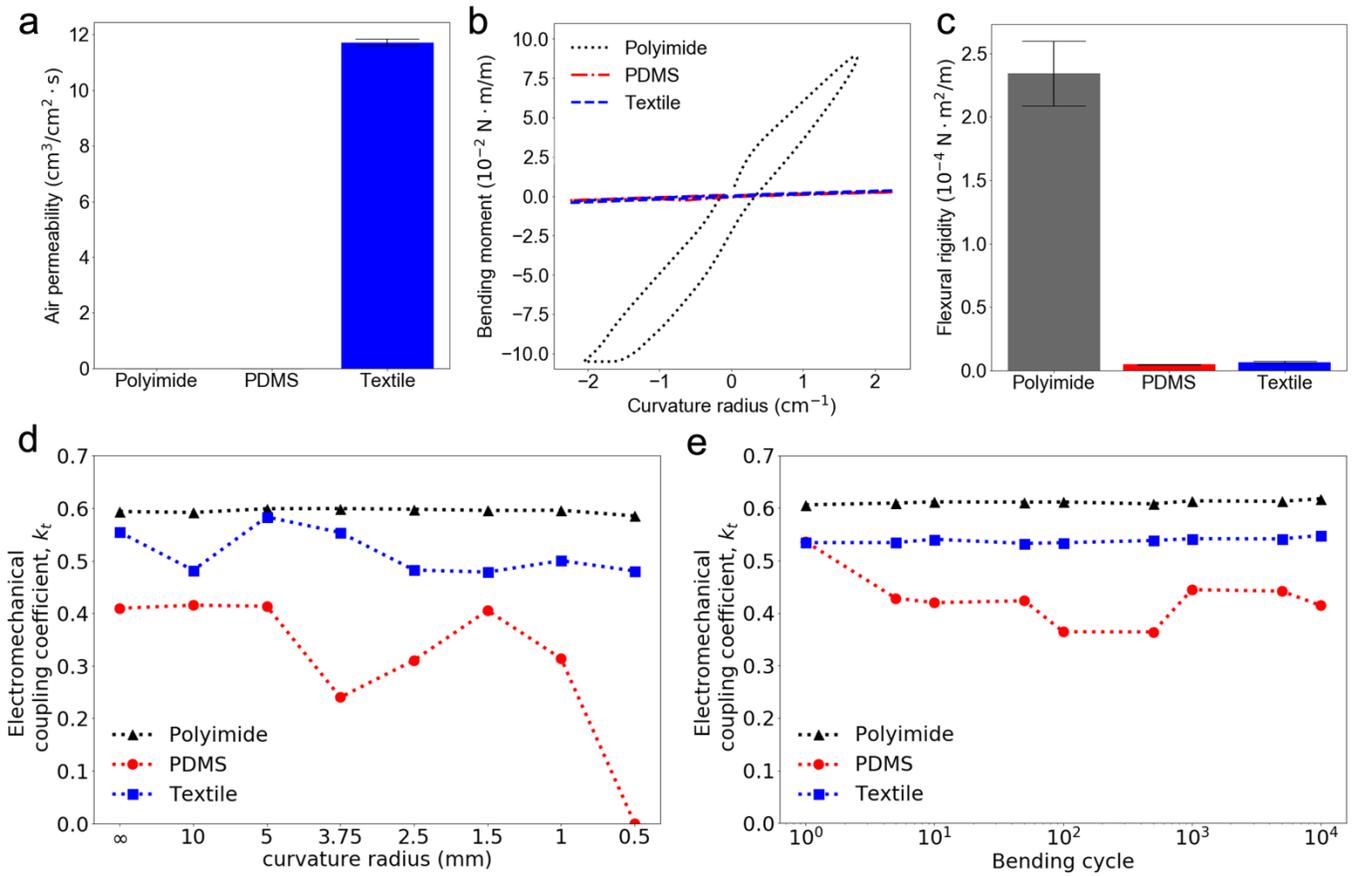

Figure 2. Wearability evaluations of the textile-based ultrasound imaging probe. (a) Frazier air permeabilities of the polyimide-based, PDMS-based, and textile-based probes. (b) Examples of the bending moments obtained by the bending test. (c) Flexural rigidities of the probes calculated from the slopes of the bending moments. (d) Evaluation results of the stabilities against the large deformations. Electromechanical coupling coefficients $k_t$ of the center element were measured after the probe bending with various diameters. (e) Evaluation results of the stabilities against the repeated deformations. $k_t$ of the center element was measured during the 10,000 cycles of repeated bending with a curvature radius of 5 mm.

## B. Wearability

The wearability of the textile-based probe was evaluated and compared with those of the polyimide-based and PDMS-based probes. First, the Frazier air permeabilities of the probes were measured using the air permeability tester (DAP-360, Daiei Kagaku Seiki Co., Ltd.). The air permeabilities were measured three times in both directions (Fig. 2a). The average air permeability of the textile-based probe was 11.7 cm$^3$/cm$^2$·s, which was significantly higher than the polyimide-based probe (0.0 cm$^3$/cm$^2$·s) and the PDMS-based probe (0.0 cm$^3$/cm$^2$·s).

Second, the probe's flexural rigidity B and the flexural hysteresis 2HB of the probes were evaluated. B and 2HB were calculated from the bending moments measured by the pure bending tester (KES-FB2-A, Kato Tech Co., Ltd.) [31]. B was obtained from the slope of the bending moment at the curvature radius of 1.0 cm$^{-1}$. 2HB was obtained from the difference in the bending moment between the loading and unloading of the bending at the curvature radius of 1 cm$^{-1}$. The probes were placed on the equipment so that the bending line came to the center of the element array. Bending moment measurements were made three times. B and 2HB were measured for both bending directions. The details of the pure bending test are shown in Fig. S5. Fig. 2b shows the examples of the bending moments. The textile-based and PDMS-based probes showed significantly lower bending moments compared with the polyimide-based probe. Fig. 2c shows the measurement results of flexural rigidity B. The average flexural rigidity of the textile-based probe (0.066 × 10$^{-4}$ $N$ ·m$^2$/m) was significantly lower than the polyimide-based probe (2.3 × 10$^{-4}$ $N$ ·m$^2$/m), and comparable to the PDMS-based probe (0.047 × 10$^{-4}$ $N$ ·m$^2$/m). The average flexural hysteresis 2HB of the polyimide-based, PDMS-based, and textile-based probes were 3.4 × 10$^{-2}$, 0.048 × 10$^{-2}$, and 0.093 × 10$^{-2}$ $N$ ·m/m, respectively.

## C. Stability against the deformation

Large and repeated deformations occur in wearable US probes that are used for daily health monitoring. Removing probes from skin can lead to large deformations. When probes are placed around joints, repeated deformations can occur. For example, when probes are attached to neck, they bend during activities such as looking at a smartphone or engaging in conversation. To be used for long-term monitoring, the textile-based US probe needs to have stability against large and repeated deformations. First, the stability against the large deformation was evaluated by the bending test. In the bending test, the probes were wrapped around the cylinders with the textile of the common ground electrode inside (Fig. S6). The diameters of the cylinders were 20, 10, 7.5, 5.0, 3.0, 2.0, and 1.0 mm. The stabilities of the probes were evaluated by electromechanical coupling coefficients (EMCC) $k_t$ of the center element after the bending. EMCC is the measure of the conversion efficiency between electrical and acoustic energy. It is widely used to evaluate the performance of the piezoelectric transducers. Fig. 2d shows the results of the bending test. While there were small decreases in EMCCs of the textile-based and polyimide-based probes, the EMCC of the PDMS-based probe significantly decreased after the wrapping around the cylinder with a 1.0 mm diameter. The significant decrease in the EMCC of the PDMS-based probe was thought to be caused by wire disconnection when the probe was bent. This wire disconnection occurred because of the difference in stretchability between the PDMS film and the wire.

Second, the stability against the repeated deformation was evaluated by the repeated bending test (Fig. S7). The probes were bent with a 5.0 mm curvature radius up to 10,000 times at 20 rpm. As in the bending test, the stabilities of the probes were evaluated with the EMCCs of the center elements. EMCCs were measured after the 1, 5, 10, 50, 100, 500, 1000, and 10,000th bend. Fig 2e shows the results of the repeated bending test. The EMCCs of the polyimide-based, PDMS-based, and textile-based probes changed little after being bent 10,000 times. The results of the bending test and the repeated bending test showed that the fabricated textile-based probe had high stability against the large and repeated deformations and had the potential for long-term monitoring application in daily life.

### D. Imaging performance

In order to evaluate the imaging performances of the textile-based probe, pulse-echo waveforms, imaging resolutions, and the abilities to visualize the blood vessels were evaluated.

First, pulse-echo waveforms of the textile-based probe were evaluated. The probe was placed on the top surface of a 10 mm thick agar gel with a concentration of 3.0 wt% (Fig. 3a). The US waves were transmitted from the elements, one element at a time, in sequence. The reflected waves from the back surface of the gel were measured by the same elements as the transmissions. Fig. 3b shows a representative example of the pulse-echo waveform and the frequency spectrum. The average -6 dB wave width of the textile-based probe was 0.69 $\mu$s, which corresponds to 1.0 mm at a sound speed of 1500 m/s. The average -6 dB bandwidth was 3.7 MHz. The wave width and bandwidth of the textile-based probe were comparable to those of the commercially available linear US probe (Fig. S8). Fig. 3c shows the relative pulse-echo wave amplitudes of the elements. The result showed that the US waves were successfully transmitted and received by all elements. On the other hand, there was a large variation in the sensitivity of the elements. The cause of the low amplitudes in some elements seemed to be poor contact between the elements and the agar gel. This result was because the probe was not pressed against or adhered to the agar gel.

Second, the imaging resolutions of the textile-based probe were evaluated with a wire phantom (N-365, Kyoto Kagaku Co.). Strings with a diameter of 0.1 mm were located in 10 mm intervals (Fig. 3d). The sound speed of the phantom was 1432 m/s, and the attenuation coefficient was 0.59 dB/cm/MHz, which were close to those of human tissue. The probe was attached to the phantom through an adhesive hydrogel sheet with a thickness of 0.7 mm. As shown in Fig. S9, the hydrogel sheet was attached to the part of the probe directly below the element array. The B-mode image was obtained by the radio frequency (RF) data acquisition and the image reconstruction. In order to acquire the RF data with a high signal to noise ratio (SNR), the synthetic aperture (SA) method using Hadamard encoding was used [32], [33]. In order to improve the B-mode image quality, the image reconstructed by the delay-and-sum (DAS) method [27] was weighted with a generalized coherence factor (GCF) [34]. The details of the RF data acquisition and the image reconstruction are provided in the methods section. Fig. 3e shows the B-mode image of the wire phantom obtained by the textile-based probe with a resolution of 20 pixels/mm. All strings up to 40 mm depth were successfully visualized. The blurred points on the left and right sides of the strings are the artifacts due to the grating lobes. Grating lobes occur when the element pitch is greater than half a wavelength. Fig. 3f shows the lateral and axial imaging resolutions of the textile-based probe. The full-width half-maxima (FWHMs) of the strings in the B-mode image were measured to evaluate the imaging resolutions. The axial FWHM was approximately 0.6 mm regardless of the depth. The lateral FWHM increased with depth, from 0.3 mm at 10 mm depth to 1.1 mm at 40 mm depth. The low lateral resolution in the deep area comes from the low f-number [35]. A comparison of the imaging resolutions of the textile-based probe with the commercially available linear probe is shown in Fig. S10.

Third, the ability of the textile-based probe to visualize the blood vessels was evaluated with a blood vessel phantom (Model ATS 524, CIRS). In the blood vessel phantom, a simulated blood vessel with a diameter of 6 mm was located at a center depth of 15 mm. The sound speed was 1450 m/s, and the attenuation coefficient was 0.5 dB/cm/MHz. The B-mode imaging was performed for longitudinal and transverse planes of the blood vessel (Fig. 3g). As in the wire phantom imaging, the textile-based probe was attached to the phantom through the adhesive hydrogel sheet. Figs. 3h and 3g show the reconstructed blood vessel images in the longitudinal and transverse planes, respectively. In the longitudinal plane image, the upper and lower walls of the blood vessel were successfully visualized. In the transverse plane image, the circular wall of the blood vessel was successfully visualized. These results suggested the possibility of visualizing the morphology of the blood vessels with the fabricated textile-based probe.

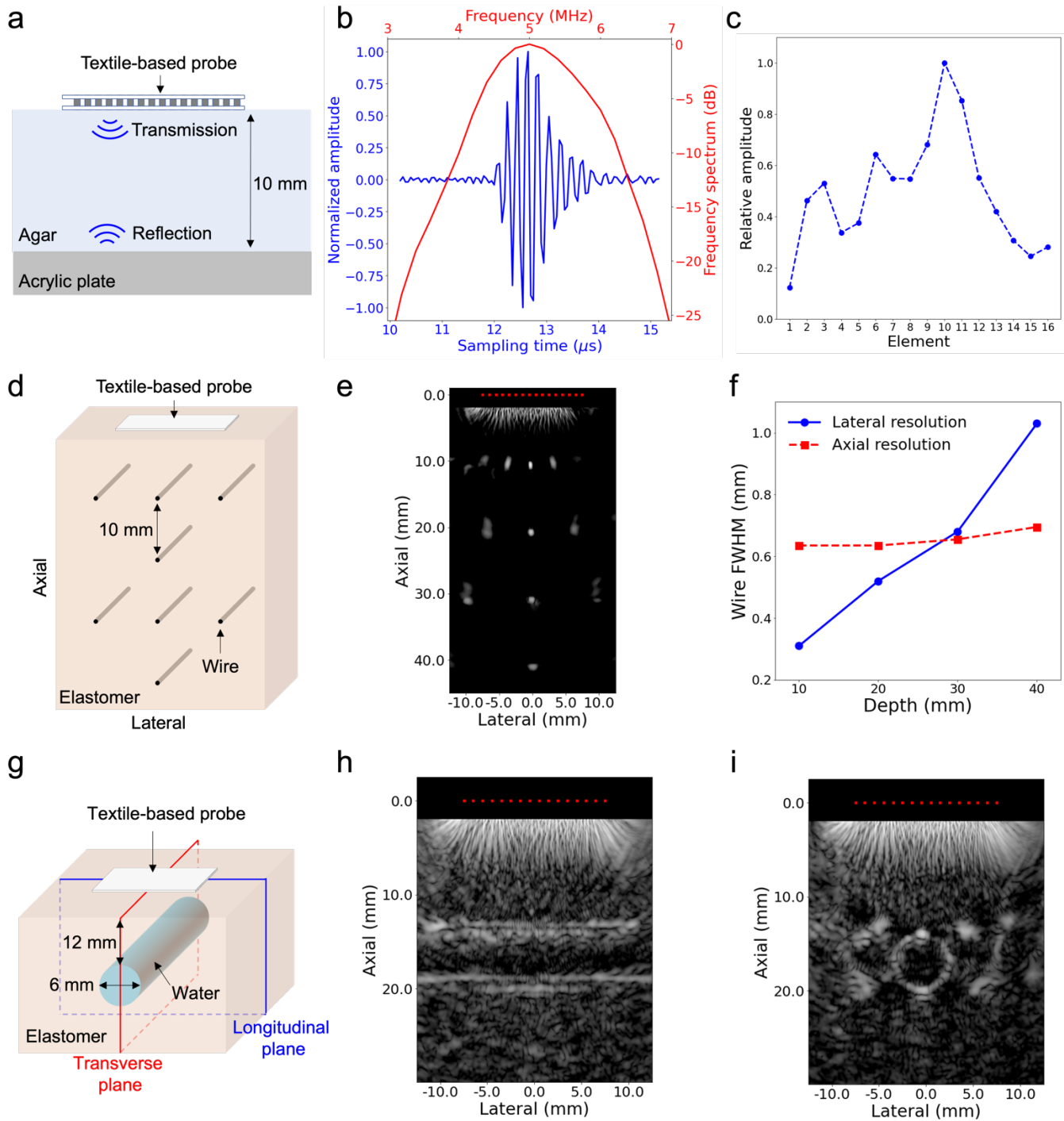

Figure 3. The imaging performance of the textile-based probe. (a) Schematics of the pulse-echo waveform measurement setup. (b) A representative example of the pulse-echo waveform (blue) and the frequency spectrum (red). (c) Relative amplitudes of the pulse-echo waves for the elements. (d) Schematics of the imaging resolution measurement with a wire phantom. (e) A reconstructed image of the wire phantom. The red squares represent the locations of the elements. The dynamic range is 40 dB. (f) Lateral and axial resolutions at each wire depth. The resolutions were evaluated by the FWHMs of the wires in the reconstructed image. (g) Schematics of the blood vessel phantom imaging. (h) (i) Reconstructed images of the blood vessel phantom in longitudinal and transverse planes, respectively. The dynamic range is 80 dB.

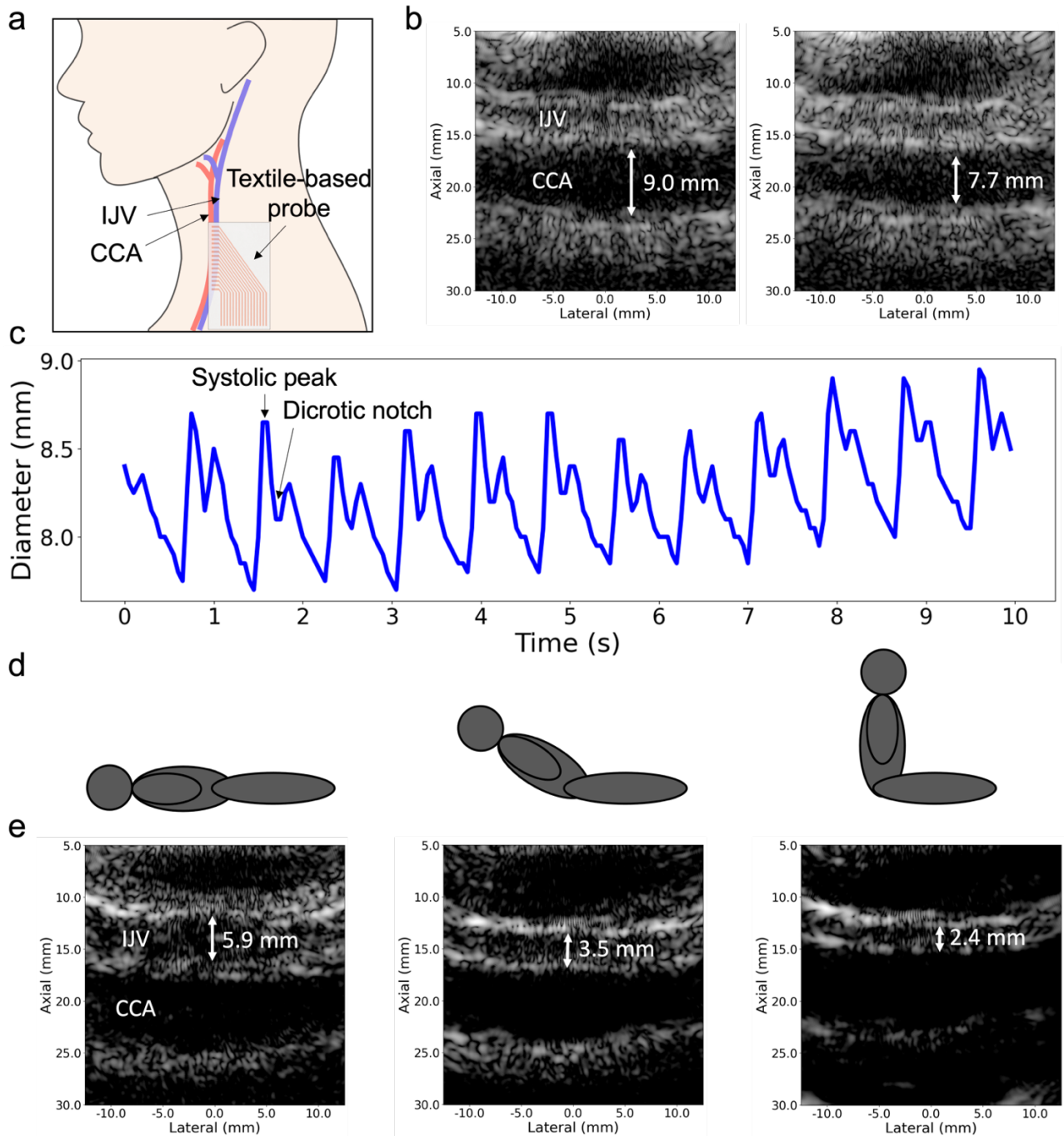

Figure 4. Monitoring of the human neck vessels using textile-based probe. (a) Schematics of the experimental setup. The probe was attached so that the IJV and CCA are included in the FOV. (b) The frames when the CCA was most dilated and constricted. Dynamic range is 40 dB. The diameter change with pulsation was successfully visualized. (c) Time-series changes in the CCA diameter for 10 s. characteristic features of the carotid artery waveforms such as the systolic peaks and the dicrotic notches could be observed. (d) Three different body postures in the IJV diameter measurements. Upper body of the subject was elevated 0, 30, 90 degrees from the supine position. (e) The US images obtained with three different body postures. Dynamic range is 30 dB. The IJV diameter changes according to the body postures could be detected.

## Monitoring of human neck blood vessels

In order to evaluate the possibility of using the textile-based probe for clinical monitoring, continuous imaging of the internal jugular vein (IJV) and the common carotid artery (CCA) was performed. The textile-based probe was attached to the subject's neck so that the longitudinal planes of the IJV and CCA could be seen (Fig. 4a). As in the phantom imaging, the probe was attached to the skin via the adhesive hydrogel sheet. The imaging was performed for 10 s at 20 fps. This experiment was approved by the University of Tokyo Ethics Committee for human studies (Approval Number: KE21-63). Fig. 4b shows the frames when the CCA was dilated and constricted. Fig. 4c shows the time-series changes in the CCA diameter. These results showed that the CCA pulsation can be monitored. Because the change in arterial diameter is reduced in various conditions such as hypercholesterolemia and congestive heart failure [36], monitoring of the CCA pulsation with the textile-based probe will enable the early detection of these diseases.

In order to demonstrate that changes in IJV diameter can be monitored with the textile-based probe, the B-mode imaging was performed with three different body postures with the upper body elevated 0 30, and 90 degrees from the supine position (Fig. 4d). The location of the probe attachment did not change in the three different postures. Fig. 4e shows the images for each body posture. The diameters of the IJV with the body posture of 0, 30, and 90 degrees were 5.9, 3.5, and 2.4 mm, respectively. This result shows that the textile-based probe can detect IJV diameter changes. Since the IJV diameter is correlated with circulating blood volume [37], continuous monitoring with the textile-based probe may allow early detection of dehydration. The advantage of using flexible probes for the IJV monitoring is that, in addition to continuous monitoring, the probes do not need to be pressed against the body. Since veins have low blood pressure and are easily contracted by pressing, the accurate measurement of the vein vessel diameters was difficult with conventional hand-held probes (Fig. S11).

### III. DISCUSSION

The textile-based US imaging probe fabricated in this study showed high flexibility, breathability, and stability against the deformations. Furthermore, the ability to obtain clinical information from the IJV and CCA was demonstrated. These results demonstrated the feasibility of early detection of the diseases by continuous monitoring with textile-based US probes.

Textiles were considered unsuitable for the US probe's substrate because they contain significant amount of air inside, which reflect the US waves (Fig. S1). However, in this study, the textile-based probe showed a pulse-echo wave amplitude comparable to the conventional linear probe and a clear visualization of the phantoms and human neck blood vessels. The textile-based probe worked correctly because the copper plating solutions and solders were soaked into the textile and filled the air gaps in the textiles. As shown in Fig. S2, the air gaps between the polyester fibers and the threads were filled with copper and solder, respectively.

Compared to plastic and elastomer films, the advantages of using textiles as US probe's substrates was not only the high flexibility and breathability, but also the great stability against deformations. As shown in Fig. 2d, the metal wires formed on stretchable substrates broke due to the significant difference in stretchability between PDMS films and copper wires. Hu *et al.* demonstrated that the serpentine-shaped geometry can improve the stretchability of metal wires [20]. However, serpentine-shaped wires limit the arrangement density of US elements, which leads to the decrease in US image quality. On the other hand, the copper wires formed on the polyester fibers in e-textiles did not break because polyester fibers had low stretchability. The reason why the e-textiles could deform even with low stretchable fibers was that their structure changed due to the rearrangement of fibers. Therefore, textile-based probes do not need to use serpentine-shaped wires, which allows for high density arrangement of US elements.

In US imaging experiments, adhesive hydrogel sheets were used to adhere the probe to the imaging objects. Although the region where the hydrogel sheet was attached was limited around the element array, the hydrogel sheet might reduce the breathability of the probe. One possible way to ensure close contact without the hydrogel sheet is to use skin-tight textiles. Skin-tight textiles are used for wearable ECG monitoring devices to provide a gel-less contact between the electrode and the skin [38]. On the other hand, the structure of the hydrogel is porous, allowing it to absorb the sweat [39]. Additionally, the hydrogel contains a large amount of water, allowing the penetration of gasses such as oxide and carbon dioxide through water [40]. Thus, the stress on the skin may be sufficiently small, even with the hydrogel sheet.

The image quality of the textile-based probe was low compared to that of the commercially available US probes. The main cause of the poor image quality was the grating lobes came from the large element pitch (1.0 mm) [27]. Note that the grating lobes do not occur when the element pitch is lower than half-wavelength (0.15 mm). The reason for using the large element pitch compared to half-wavelength was that there was a possibility of a short circuit between adjacent wires in the e-textiles when a narrow wiring pitch was used. Using a densely woven textiles with finer threads may enable the development of a textile-based probe with narrower element pitch.

The textile-based probe was designed without a backing layer to maintain flexibility and breathability, which is typically attached to the backside of the elements to absorb US waves. The lack of a backing layer can lead to prolonged element vibration and pulse-wave elongation. However, despite the absence of a backing layer, the pulse-echo wave width of the textile-based probe remained comparable to that of commercially available linear probes (Fig. S8). This result can be attributed to the absorption of US waves by the top e-textile. As shown in Fig. S12, the US waves transmitted from the top side of the probe had significantly smaller amplitudes than those transmitted from the bottom side. The absorption by the top textile may come from the smaller width of the signal electrodes relative to that of the common ground electrode. There is a possibility that the air gaps in the textiles at the electrodes may not be completely filled with copper if the electrode widths are too narrow. The presence of small air gaps remained in the textile seemed to have contributed to the US wave absorption.

Since the textile-based probe was attached to the skin, its position could not be adjusted during the imaging. A large field of view (FOV) is required to ensure that the imaging target is included in the FOV, despite the attachment position errors. Moreover, since the attachment position errors occur in azimuth and elevation directions, a two-dimensional element array is required to enable the extraction of a desired imaging plane from a three-dimensional FOV.

## IV. CONCLUSION

In this study, we have successfully fabricated the textile-based ultrasound imaging probe for long-term health monitoring, which had high breathability, flexibility, and stability against deformations. Experimental results showed that the probe can clearly visualize the blood vessels and monitor the clinical information. This study demonstrated the feasibility of early detection of the diseases by the long-term US imaging with textile-based probes in daily life.

## V. METHODS

### A. Electromechanical coupling coefficient measurement

The following processes measured the EMCC of the center element. First, the impedance spectrum of the element was obtained using an impedance analyzer (4294A, Agilent Technologies, Inc.). Second, the local minimum and maximum of the absolute impedance were extracted as the resonant frequency $f_r$ and anti-resonant frequency $f_a$, respectively (Fig. S13). Finally, the thickness mode EMCC $k_t$ was calculated from $f_r$ and $f_a$ by the following equation [41].

$$k_t = \sqrt{\frac{\pi}{2}\frac{f_r}{f_a}\tan\left(\frac{\pi}{2}\frac{f_a - f_r}{f_a}\right)} \quad (2)$$

### B. RF data acquisition

For all the experiments in the imaging performance evaluation and human blood vessel monitoring, US wave transmission and RF data acquisition were performed using a US signal acquisition system (V1, Verasonics, Inc.). For the US wave transmissions, the elements were excited with a two-cycle electric signal at 5 MHz at an amplitude of 50 V. The backscattered US waves received by the elements were recorded at a sampling rate of 20 MHz for 1,536 times.

For the US imaging, we used Hadamard-encoded SA to obtain the RF data. In the SA without Hadamard encoding, US pulse waves are transmitted from all 16 elements in the sequence, one element at a time. For every transmission, backscattered waves received by all elements are recorded. The resulting RF data size is 16 (number of transmissions) × 16 (number of receiving elements) × 1,536 (number of sampling times). SA is suitable for the deformable US probes because it does not require the elements' positions at the time of RF data acquisition. However, SA has the disadvantage of low SNR due to the single-element transmissions. In order to overcome the low SNR of the SA, Hadamard encoding was used. In the Hadamard encoded SA, all elements were excited with apodizations of +1 or -1. The apodization of the $j$th element at $i$th transmission was $H_{16}(i,j)$. Where $H_{16}$ is the Hadamard matrix of order 16 [42]. After the acquisition, the RF data undergo decoding by multiplying with the Hadamard inverse matrix. The resulting RF data are equivalent to those obtained with single-element transmissions, while the SNR is significantly improved. Figs. S14 and S15 show the image quality comparison between with and without the Hadamard encoding. Although there are other methods to overcome the low SNR of SA, such as virtual sources [43] and plane waves [44], these methods are difficult apply for flexible probes because they require the element array geometry.

### C. Image reconstruction

The image reconstruction was performed as follows. First, the RF data were filtered with a band-pass filter to remove the noises whose frequencies are different from the transmitting signals. The cutoff frequencies were 3.0 and 7.0 MHz. Second, time gain compensation with a coefficient of 0.5 dB/cm/MHz was applied to the RF data to enhance the attenuated signals from deep tissues. Third, for the DAS reconstruction, the delayed signals were collected from the RF data. The delayed signals $d$ are the three-dimensional data, each of which is as follows.

$$d(tx, rx, p) = y(tx, rx, t) \quad (3)$$

where $tx$ and $rx$ are the number of transmitting and receiving elements, respectively. $p$ is the index of an imaging pixel. $y$ is the RF data. $t$ is the time for the US wave to propagate from the transmitting element to the receiving element through the imaging point, which was calculated by the following equation.

$$t = \frac{|\vec{r_p} - \vec{r_{tx}}| + |\vec{r_{rx}} - \vec{r_p}|}{c} \quad (4)$$

where $\vec{r_p}$, $\vec{r_{tx}}$, and $\vec{r_{rx}}$ are the coordinates of the imaging point, the transmitting element, and the receiving element, respectively. $c$ is the sound speed. In this study, the coordinates of the elements used for the image reconstructions were determined by assuming that the elements were aligned in a straight line. This was because the surface shapes of the phantoms were flat, and the surface shape of the human neck was almost linear along the longitudinal axis. In order to apply the textile-based probe for the imaging with unknown surface shapes, elements' positions need to be estimated [45]. Next, the B-mode image was obtained by summing the delayed signals. The brightness of each pixel in the image $I_{DAS}$ was as follows:

$$I_{DAS}(p) = \sum_{tx=1}^{16} \sum_{rx=1}^{16} d(tx, rx) \tag{5}$$

Finally, the image was weighted with GCF to enhance the image quality. The GCF was the ratio of the spectral energy of the delayed signals in the low-frequency region to that of the total region. The GCF of each pixel was calculated as follows:

$$GCF(p) = \frac{\sum_{k=-M_0}^{M_0} \sum_{l=-M_0}^{M_0} |\mathcal{F}[d](k,l)|^2}{\sum_{k=-8}^{7} \sum_{l=-8}^{7} |\mathcal{F}[d](k,l)|^2} \tag{6}$$

where $\mathcal{F}[d]$ is the two-dimensional Fourier transform of the delayed signals. $k$ and $l$ are the first and second indexes of $\mathcal{F}[d]$. $M_0$ is the parameter to determine the low-frequency region. In this study, $M_0$ was set to 1. Since the GCF weighting suppresses the incoherent signals, noise and grating lobes were reduced. A comparison of the reconstructed images with and without the GCF weighting is shown in Figs. S14 and S15.

### D. Imaging resolution evaluation

The imaging resolutions of the textile-based probe at each depth were measured from the reconstructed image of the wire phantom by the following processes. First, regions with a size of 5 mm × 5 mm around the wires were extracted from the reconstructed images. The center positions of the regions were 0 mm (lateral) and 10, 20, 30, and 40 mm (axial). Then, the pixel with maximum brightness was found in each region. The row and column with the same axial and lateral index as the pixel of maximum brightness were extracted, respectively. The FWHMs of the row and column were measured as the lateral and the axial resolutions. Fig. S16 shows an example of imaging resolution evaluation.

### E. Blood vessel diameter measurement

The CCA and IJV diameters were measured in the monitoring of the human neck blood vessels. In the brightness distribution of the reconstructed image at the lateral coordinate of 0 mm, the first peak from the body surface was taken as the upper wall of the IJV, the second peak was taken as the lower wall of the IJV and the upper wall of the CCA, and the third wall was taken as the lower wall of the CCA. Thus, the distance between the first and the second peaks was defined as the IJV diameter, and the distance between the second and third peaks was defined as the CCA diameter. Before diameter measurement, the images were filtered by a Gaussian filter with a sigma of 2.5 mm in the lateral direction and 0.5 mm in the axial direction to reduce the effect of speckle noise. An example of the diameter measurement procedure is shown in Fig. S1s7.